\documentclass[a4paper, amsfonts, amssymb, amsmath, preprint, showkeys, nofootinbib, twoside]{revtex4-1}
\usepackage[english]{babel}
\usepackage[utf8]{inputenc}
\usepackage[colorinlistoftodos, color=green!40, prependcaption]{todonotes}
\usepackage{soul}
\usepackage{amsthm}
\usepackage{mathtools}
\usepackage{physics}
\usepackage{xcolor}
\usepackage{graphicx}
\usepackage[left=23mm,right=13mm,top=35mm,columnsep=15pt]{geometry} 
\usepackage{adjustbox}
\usepackage{placeins}
\usepackage[T1]{fontenc}
\usepackage{lipsum}
\usepackage{csquotes}
\usepackage[pdftex, pdftitle={Article}, pdfauthor={Author}]{hyperref} 
\usepackage{algorithm}
\usepackage{algpseudocode}
\begin{document}
\title{Complete reconstruction of the space-time dynamics in a Kerr-lens mode-locked laser}

\author{$\text{Idan Parshani}^1$, $\text{Leon Bello}^{1,2}$, $\text{Mallachi-Elia Meller}^1$, $\text{Avi Pe'er}^1$}
    \email[Correspondence email address: ]{avi.peer@biu.ac.il}
     \affiliation{1 Department of Physics and BINA Institute of Nanotechnology, Bar-Ilan University, Ramat-Gan 52900, Israel}
     \affiliation{2 Department of Electrical and Computer Engineering, Princeton University, Princeton, NJ}
    
\date{\today} 

\begin{abstract}
We present a complete numerical analysis and simulation of the full spatio-temporal dynamics of Kerr-lens mode-locking (KLM) in a laser on all time-scales. The KLM dynamics, which is the workhorse mechanism for generating ultrashort pulses, relies on the intricate coupling between the spatial nonlinear evolution due to self focusing and the temporal nonlinear compression due to self-phase modulation (SPM) and dispersion. Our numerical tool emulates the dynamical evolution of the optical field in the cavity on all time scales: the fast time scale of the pulse envelope within a single round trip, and the slow time-scale between one round-trip to the next. We employ a nonlinear ABCD formalism that fully handles all relevant effects in the laser, namely - self focusing and diffraction, dispersion and SPM, space-dependent loss and gain saturation. We confirm the validity of our model by reproducing the pulse-formation in KLM in all aspects: The evolution of the pulse energy, duration, and gain is observed during the entire cavity buildup (from spontaneous noise to steady state), demonstrating the nonlinear mode competition in full, as well as the dependence of the final pulse in steady state on the interplay between gain bandwidth, dispersion and self-phase modulation. The direct observation of the nonlinear space-time evolution of the pulse is a key enabler to analyse and optimize the KLM operation, as well as to explore new nonlinear space-time phenomena. 
\end{abstract}


\maketitle

\section{Introduction}
Kerr-lens mode-locking (KLM) is the state-of-the-art technique for generating ultrafast pulses, central to the field of ultrafast lasers and extensively used \cite{herrmann_theory_1994, t_brabec_kerr_1992, t_brabec_hard-aperture_1993}. KLM relies on the highly non-linear interplay between the spatial and temporal properties of the field inside the laser cavity that is challenging to analyze fully \cite{kurtner_mode-locking_1998, matsko_mode-locked_2011, haus_mode-locking_2000} and in many cases requires approximations that exclude some of the intricacies of these oscillators. As a result, the design of KLM-based oscillators often relies strongly on intuition and trial-error experience. 
Standard methods of analysis usually focus on the steady-state of the field \cite{MAGNI1993348,coen_modeling_2012, salin_mode_1991, yoo_byung_duk_numerical_2005,Juang1997,t_brabec_kerr_1992,Cerullo1994,Henrich1997}, but cannot describe the dynamical evolution towards this steady state. These methods make different approximations to separate the spatial and temporal dynamics, which although effective for understanding the basic operation of KLM lasers, can miss important dynamics in the interplay between the spatial and temporal profiles of the field \cite{parshani_diffractive_2021}. 

Here we introduce a numerical simulation algorithm, along with an open-source MATLAB program that calculates from first principles the complete dynamics of the spatio-temporal field profile in a KLM oscillator for a wide range of operation regimes. Our method is able to reproduce the evolution of the field in both space and time - starting from an initial noise-seed up to steady-state pulses. The only assumption being made is that the spatial beam of the laser is a single mode, well approximated by a Gaussian profile. This is a practically universal regime of operation in KLM lasers, which is strongly driven to a single spatial mode by the interaction between the self-focusing Kerr-lensing and the diffraction through the effective aperture in the cavity \cite{t_brabec_kerr_1992, t_brabec_hard-aperture_1993}.

Our paper is structured as follows: Section \ref{sec:results} includes the major validations, where the simulation reconstructs fundamental properties of KLM oscillators with both qualitative and quantitative agreement - soliton formation from spontaneous noise and its dependence on the laser parameters, such as gain bandwidth, gain saturation, SPM, linear and nonlinear dispersion, all from first principles. This section also shows how different pulse properties depend delicately on the spatial cavity parameters and how we can use the full simulation of a KLM laser to assist cavity design. Finally, Section \ref{sec:methods} goes into a detailed description of the paradigm of our numerical simulation and the algorithm.

\section{Simulating the dynamical formation of the pulse}
\label{sec:results}

Before we delving into the intricacies of  the algorithm of the numerical simulation, let us demonstrate the performance of the simulation by reconstructing the complete dynamical evolution of the pulse, from the initial spontaneous noise seed, through the entire cavity buildup to the final steady state. We will show how the spatial mode of the cavity is formed in steady state by self-focusing and how the soliton pulse is dictated by the interplay between gain-bandwidth, dispersion, and SPM.

\begin{figure}[h!]
    \centering
    \includegraphics[width=\textwidth]{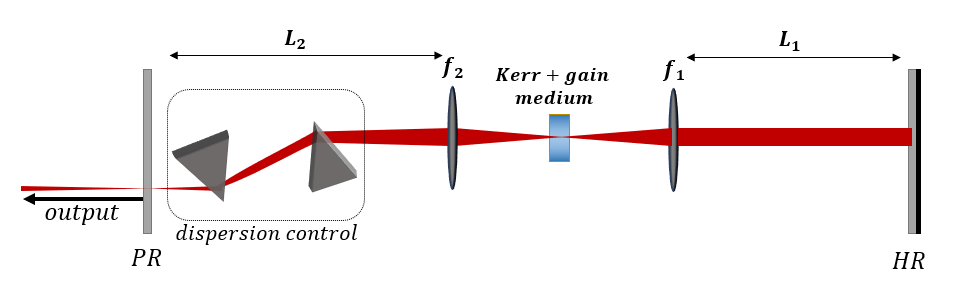}
    \caption{The cavity layout of a standard KLM laser in a linear cavity: The cavity focus is formed by two spherical mirrors / lenses), where the nonlinear Kerr medium (and gain) is placed. Two cavity arms of lengths $L_1,L_2$ exist around the focus. One of the arms includes components for tuned dispersion control (e.g. a double-prism).}
    \label{fig:sketch.}
\end{figure}

\begin{figure}[h!]
    \centering
    \includegraphics[width=\textwidth]{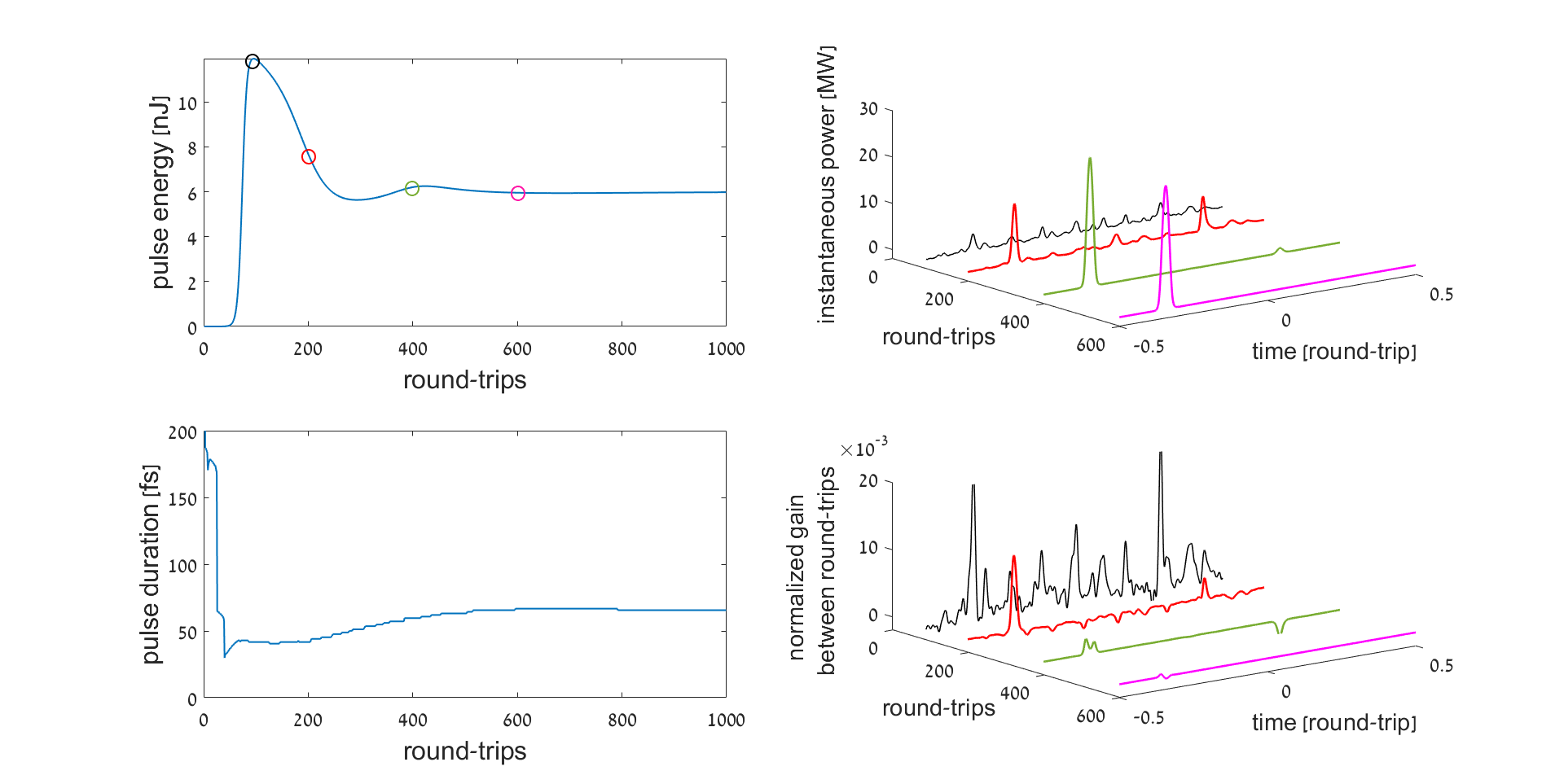}
    \caption{Evolution of the laser field during the entire oscillation buildup: (a) The cavity (pulse) energy and (b) the pulse-duration (FWHM) as a function of round-trip number (slow time scale). To observe the evolution of the pulse field withi the cavity round-trip, we sampled four specific round-trips, as shown in (c) for the instantaneous power and (d) for the instantaneous gain as a function of time within a single round-trip (fast time scale). The colors of the graphs in c,d correspond to the circles on the energy graph of (a).}
    \label{fig:dynamics}
\end{figure}

The simulation assumes the standard linear cavity layout of figure \ref{fig:sketch.}, and the evolution of the intra-cavity field during the laser buildup is highlighted in figure \ref{fig:dynamics}. The cavity (pulse) energy (figure \ref{fig:dynamics}a) and the pulse-duration (FWHM, \ref{fig:dynamics}b) are shown as a function of round-trip number (the slow time scale), where three stages of evolution are evident \textbf{(I)} the initial exponential growth, \textbf{(II)} the nonlinear mode competition and \textbf{(III)} the final steady-state. To understand the dynamical operation during each stage, we sampled four specific round-trips from the different stages, which are shown in figure \ref{fig:dynamics}c (instantaneous power) and \ref{fig:dynamics}d (instantaneous power gain) as a function of time within a single round-trip (fast time scale). The colors of the graphs in figures \ref{fig:dynamics}c,d correspond to the circles on the energy graph of figure \ref{fig:dynamics}a. The field at the end of the initial exponential growth (black) is highly noisy and multi-pulsed, with noisy, yet positive gain at all times. This picture changes during the mode competition (red and green). We observe a number of pulses competing for the gain, before all pulses, except one, diminish. During that phase, we also observe variations in the shape of the main pulse. Finally, the purple graph shows the stable pulse near steady state.

\begin{figure}[h!]
    \centering
    \includegraphics[width=0.9\textwidth]{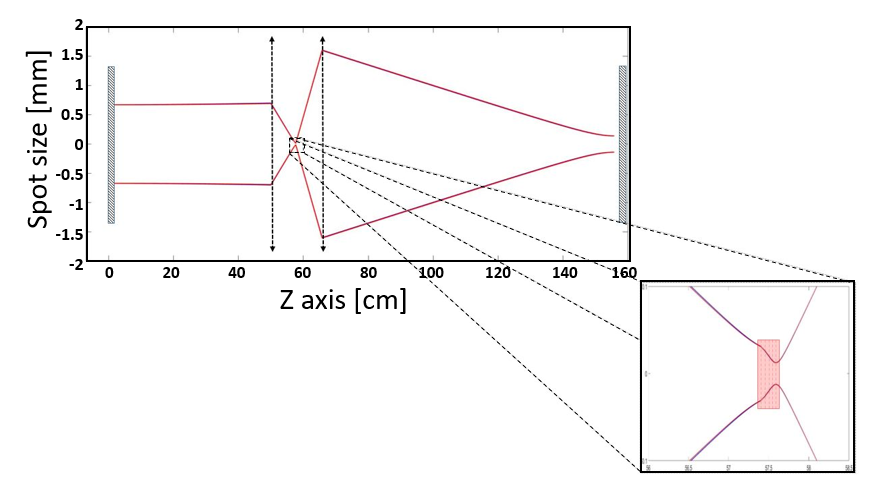}
    \caption{\textbf{(a)} The temporal profile of the pulse. The initial noise seed quickly converges into a steady-state pulse, where the beam waist varies in time, completely correlated with the temporal profile, and the spectral phase is flat, as expected from a transform-limited pulse. \textbf{(b)} The beam propagation through the cavity in a representative round-trip. The non-linear focus bends the beam back into stable operation, creating an effective saturable absorber -- low-power pulses are unstable, while high-power pulses are focused back into stability.}
    \label{fig:noise_to_pulse}
\end{figure}

Our simulation is fully spatio-temporal, and calculates the beam propagation through the cavity in space as well as in time. Figure \ref{fig:noise_to_pulse} shows the propagation of the beam through a single run of the cavity. The core function of the Kerr-lens is shown in the inset, where the simulation captures the lensing effect that creates an effective "wave-guide" that stabilizes the cavity and counteracts the diffraction losses.

\begin{figure}[h!]
    \centering
    \includegraphics[width=0.7\textwidth]{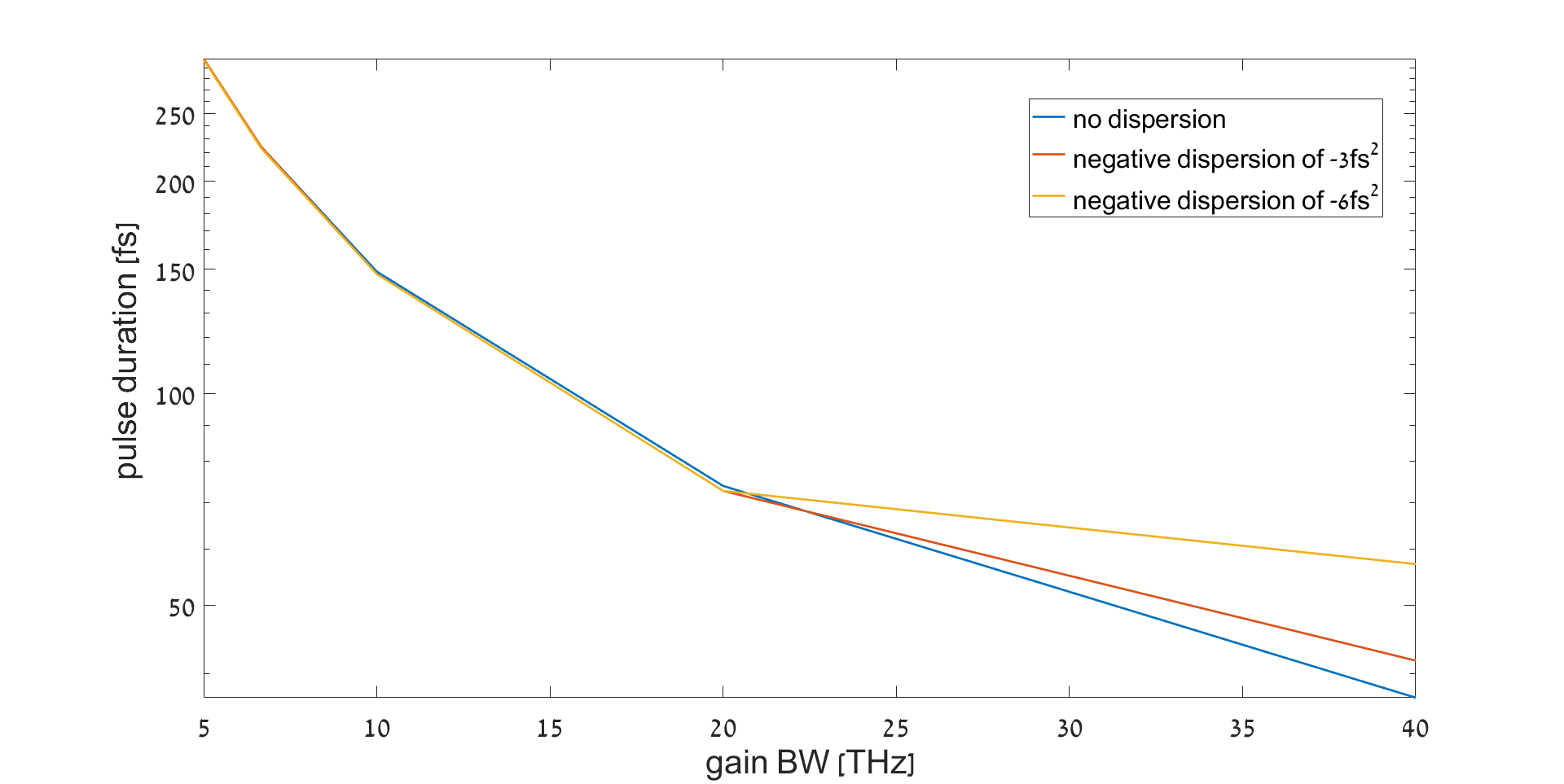}
    \caption{The temporal width of the pulse in time as function of gain bandwidth. As expected, the pulse becomes shorter and shorter as bandwidth increases. Furthermore, we see that dispersion broadens the pulse, with the high dispersion configuration almost an order of magnitude bigger than the low dispersion one.}
    \label{fig:BW}
\end{figure}

The temporal width of the pulse is dictated by the gain bandwidth and the chromatic dispersion (GVD). The gain-bandwidth sets a lower bound on the pulse duration (by Fourier uncertainty), and the dispersion distorts the temporal shape of the pulse (chirping), which affects shorter pulses more severely than longer pulses \cite{Proctor:93}. Thus, the net GVD of the cavity sets another lower bound on the pulse-duration. Our simulation correctly reproduces this behaviour, as shown in figure \ref{fig:BW}, where the pulse duration follows quantitatively the gain bandwidth limit, until it reaches the dispersion limit. 

\begin{figure}[h!]
    \centering
    \includegraphics[width=\textwidth]{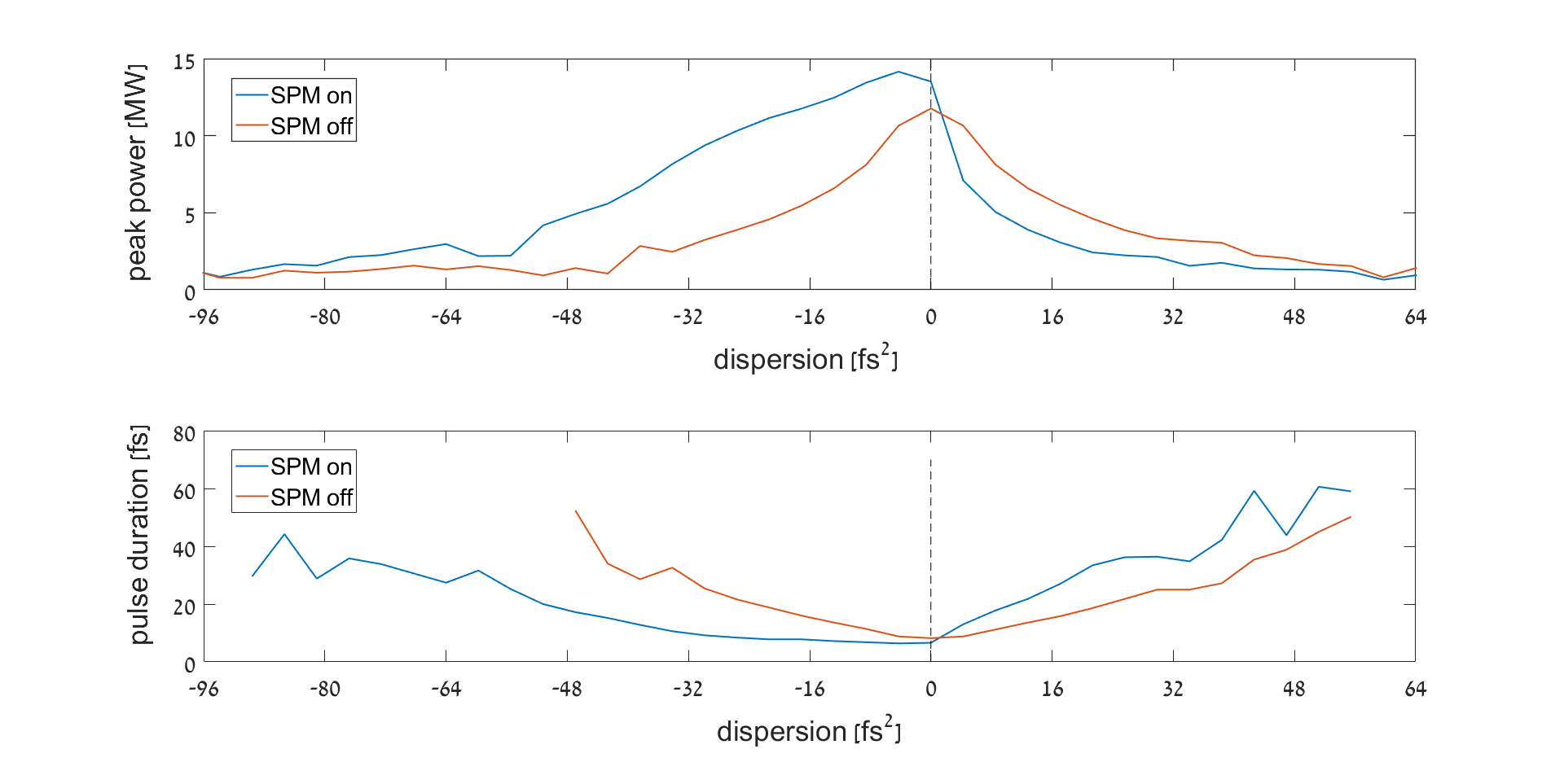}
    \caption{The width and peak power of the pulse as a function of the overall dispersion. The pulse reaches maximum power and minimum width for a specific, non-zero value, due to the dispersion induced by the non-linear Kerr-effect.}
    \label{fig:dispersion}
\end{figure}

So far, we considered only linear effects on the pulse duration due to GVD and gain bandwidth. The pulse, however, is highly affected by the nonlinear chirping due to SPM, which is critical in the soliton formation mechanism. Specifically, SPM chirping acts as another source of dispersion that shifts the optimal GVD value towards a small negative value. Surprisingly, the simulation can test this interplay between SPM and GVD in greater detail than possible in experiment: In the simulation we can simply turn SPM effects on and off and compare the pulse performance, a feat that is inherently impossible in the experiment. Figure \ref{fig:dispersion} highlights exactly this comparison: we plot the pulse duration as a function of the linear GVD in the cavity with SPM (blue) and without (orange). Indeed, the simulation shows that SPM shifts the optimal pulse formation towards the negative GVD range as expected, but also that SPM enhances the peak power at the optimal GVD value. Thus, optimal operation is achieved at some dispersion, rather than the zero dispersion point.

Beyond the validations presented above, this simulation already served two previous publications, where it was used to predict and analyse new details of KLM. Specifically, in \cite{parshani_diffractive_2021} we used the simulation to observe and explore the effective saturable-absorption mechanism of KLM, where although no actual absorption takes place, diffraction-losses over an effective aperture form the saturation loss mechanism. Using the simulation, we identified how and where exactly these losses occur, which aided us to measure this time-dependent loss for the first time. In another paper \cite{Parshani2021}, we used the simulation to show that KLM lasers can break the spatial symmetry between the forward and backward halves of the round-trip in a linear cavity. when the pump power is increased above the mode-locking threshold. This symmetry breaking allows the laser to increase the energy of the pulse. Both of these phenomena would have been impossible to predict without a full dynamical simulation, which illustrates the utility of this simulation tool. 

\section{The Algorithm}
\label{sec:methods}

The simulation assumes a set of common parameter values for the cavity configuration, gain medium and Kerr medium, that are typical for KLM experiments \cite{yefet_mode_2013, yefet_review_2013}. The values used in the simulation appear in Table \ref{tab:experimental values}.

\begin{table}[h!]
    \centering
    \begin{tabular}{c|c}
        \textbf{Parameter} & \textbf{Value} \\
        \hline
        NL refractive index - $n_2$ & $3 \times 10^{-16} {cm}^2/W$ \\
        gain medium length- $L$ & $3 mm$ \\
        short path - $L_1$ & $50 cm$ \\
        long path - $L_2$ & $90 cm$ \\
        radius or curvature mirrors - $R$ & $150 mm$ \\
        pump waist in gain medium - $w_p$ & $30 \mu m$
    \end{tabular}
    \caption{Parameter values for the cavity configuration and the Kerr medium.}
\label{tab:experimental values}
\end{table}

\begin{figure}[h!]
    \centering
    \includegraphics[width=0.9\textwidth]{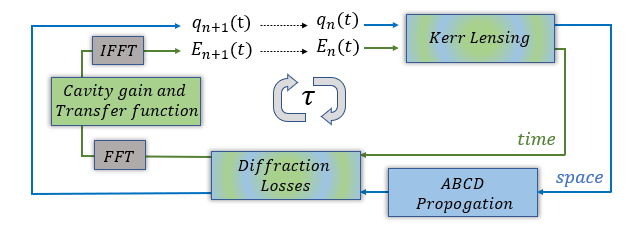}
    \caption{Flowchart of the simulation. The roundtrip time is denoted $\tau$, and includes a spatial part (outer loop - blue), which propagates the beam parameter $q(t)$ through the cavity with the ABCD formalism; and a temporal part (inner loop - green), which propagates the field in time $E(t)$ from one round-trip to the next according to the spectral transfer function of the cavity, which includes gain saturation and the spectral gain-curve, loss and chromatic dispersion (all conveniently calculated in the frequency domain with a fast Fourier transform - FFT). The two parts are coupled together via the Kerr-lensing effect and the diffraction losses, where the temporal intensity dictates the instantaneous nonlinear lens and the temporal beam parameter $q(t)$ dictates the instantaneous diffraction losses. }
    \label{fig:flowchart.}
\end{figure}

Our simulation focuses on KLM oscillators in a single-spatial mode, which is time-dependent due to the Kerr-lens. This is the usual operating regime of practically all KLM lasers since the nonlinear effect strongly drives the laser towards a single, high-intensity spatial mode. These oscillators can be thought of as a passive cavity, with an intensity-dependent lens (or lenses) that couples the temporal and spatial dynamics. When the diffraction losses are also included, the Kerr-lensing effect generates an effective saturable absorber with an instantaneous response, able to generate extremely short pulses \cite{Kartner1996, kurtner_mode-locking_1998, haus_theory_1975,c_j_chen_self-starting_1995}.

As such, the KLM oscillator can be divided conceptually into two parts - linear and non-linear. The linear part accounts for the gain spectrum and saturation, dispersion and linear loss, which can be encapsulated into a total spectral transfer function:
\begin{equation}
    E(\omega) \rightarrow g(\omega) T(\omega) E(\omega),
\end{equation}
where $g(\omega)$ is the spectral gain of the active medium and $T(\omega)$ is the transfer function of the passive elements in the cavity. The spectral phase of $T(\omega)$ reflects the chromatic dispersion and its amplitude reflects the cavity loss $\left|T\right|^2=1-Loss$. 
To take into account the gain depletion, which leads to non-linear saturation and mode competition, we reduce the gain $g(\omega)$ as the intra-cavity pulse energy $U$ increases according to the standard saturation formula [\cite{rp_photonics_sat_power}],
\begin{equation}
    g(\omega) = \frac{g_0(\omega)}{1 + U/U_{\text{sat}}},
\end{equation}
where $U_{\text{sat}}$ is the saturation energy of the gain medium, taken to be $\tau \cdot 2.6\text{W}$ for our simulation ($\tau$ is the round trip  and the saturation power of 2.6W is based on the saturation intensity of Ti:Sapphire in the literature, calculated for a 3mm long crystal with $0.25\%$ doping and $20\mu m$ beam waist for the pump). 

We now need to account for the nonlinear response of the Kerr medium, i.e. Kerr-lens and SPM, which we assume to be instantaneous. For simplicity we start with a thin Kerr medium, whose thickness $d$ is much smaller than the beam Rayleigh range $d\ll z_R$, and later generalize the result for a thick medium. The ABCD matrix of a thin Kerr medium is a simple lens with an intensity dependent focus: 
\begin{equation}
    M_{\text{KL}}(t) = 
    \begin{pmatrix}
        1 & 0 \\ -\frac{1}{f_{KL}(t)} & 1
    \end{pmatrix}
\end{equation}
where $f_{KL}$ is the time dependent Kerr-lent focus, which for a thin nonlinear medium is given by
\begin{equation}
\label{eq:4}
   \frac{1}{f_{KL}} = \frac{8 n_2 d}{\pi w(t)^2} I(t),
\end{equation}
with $n_2$ the nonlinear index of refraction, $d$ the medium length, $w(t)$ is the instantaneous beam waist and $I(t)$ the instantaneous intensity. The SPM of a thin Kerr medium is a simple phase modulation of $e^{i4n_{2}I(t)d/\pi\lambda}$.
 
This introduces a new, fast time scale $t$ into the problem. Unlike the slow laser dynamics, which occurs on a time scale of many round-trips (from one round-trip to the next), the instantaneous Kerr-lens evolves on pulse time-scale within a single cavity round-trip. As a result, we use two different time variables to describe the cavity dynamics - $n$, which enumerates the round-trip time, and a fast time variable $t$ which measures the time within each round-trip.

The beam, under the Gaussian approximation, is completely described by two variables - the instantaneous power $P_n(t)$ and Gaussian beam parameter $q_n(t)$ that reflects the beam waist $w$ and phase curvature $R$ as $\frac{1}{q}=\frac{1}{R}-i\frac{\lambda}{\pi w^2(t)}$. The local intensity is $I_n(t)=\frac{P_n(t)}{w^2(t)}$. With the instantaneous power and beam parameter we can calculate the instantaneous matrix $M_{KL}(t)$. The total ABCD matrix of the cavity round trip can now be written. For a ring cavity it is 
\begin{equation}
M_{ring}(t)=M_{cav}M_{KL}(t)
\end{equation}. 
However, for a linear cavity, the Kerr interaction occurs twice, which leads to 
\begin{equation}
M_{lin}(t)=M_{l}M^{+}_{KL}(t)M_{r}M^{-}_{KL}(t), 
\end{equation}
where $M_{l,r}$ are the linear ABCD matrices for the two halves of the cavity around the Kerr medium and $M^{\pm}_{KL}$ are the nonlinear lens matrices in the forward and backward propagation through the cavity, which are not necessarily equal. In fact, the dual interaction with the Kerr medium can break the symmetry between the forward and backwards direction, as we showed in \cite{Parshani2021}.

For a thick Kerr medium, whose thickness cannot be neglected relative to the Rayleigh range, we must also account for the propagation of the beam within the nonlinear medium. This can be approximated with a split-operator approach, where the thick medium is divided into $N_{KL}$ thin lenses that are separated by a short distance $z=d/N_{KL}$ of free propagation. The total matrix of the Kerr medium then becomes
\begin{equation}
    M_{KL}(t)=\prod_{i}{M(z)M^i_{KL}(t)},
\end{equation}
where $M(z)$ is the free space matrix and $M^i_{KL}(t)$ are the thin Kerr-lenses of equation \ref{eq:4}, each calculated according to the beam parameter at its location $q^i(t)$. In our simulation, the Kerr medium of thickness  $d=3\text{mm}$ was divided to $N_{KL}=5$ thin lenses.

\begin{figure}[h!]
    \centering
    \includegraphics[width = \textwidth]{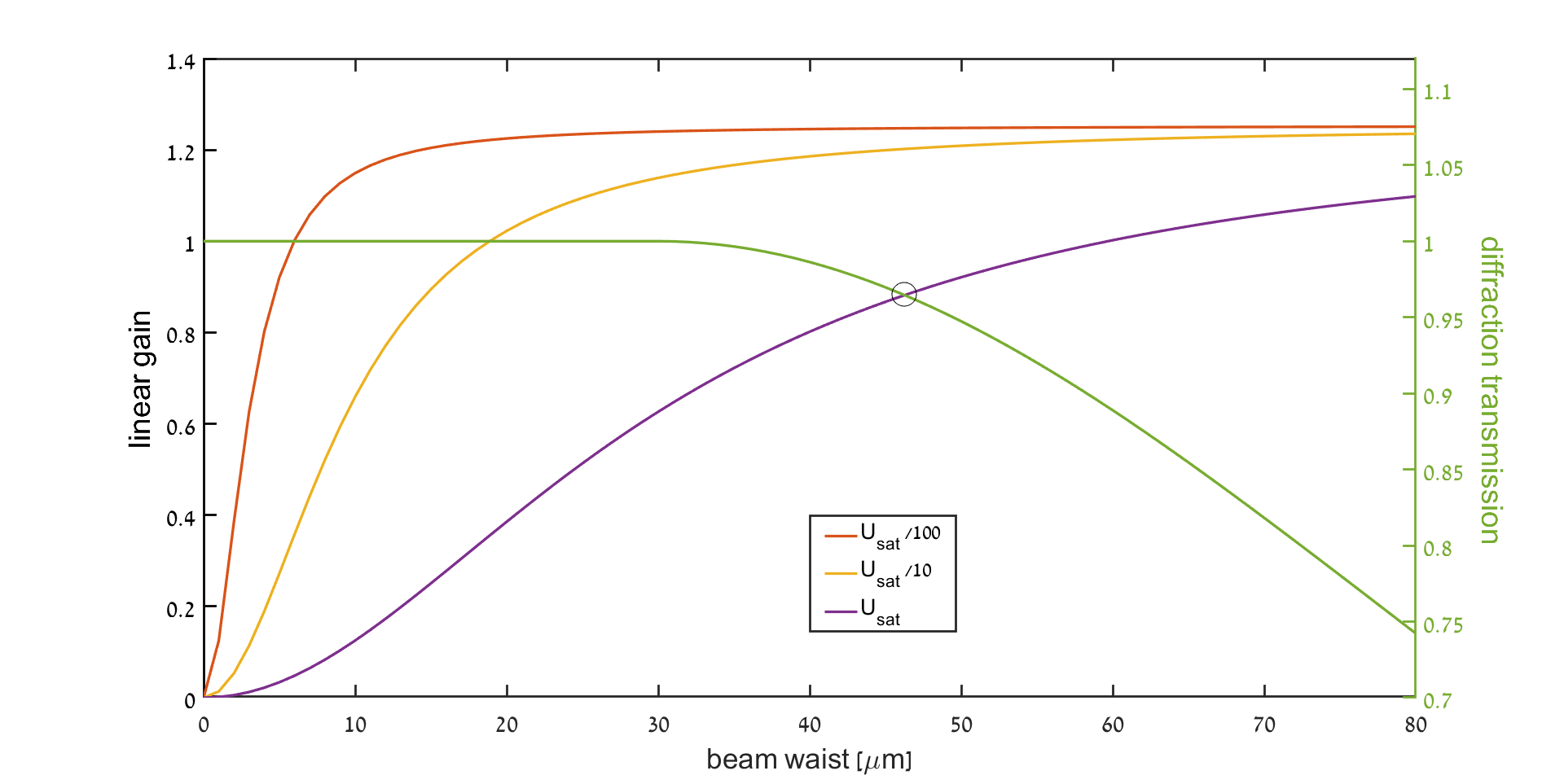}
    \caption{Modeling the diffraction losses: \textbf{Left axis} - gain as function of the beam waist for three pulse energies -  $U_{sat}/100$ (red), $U_{sat}/10$ (yellow), $U_{sat}$ (purple). The saturated gain value depends on the beam waist because for narrow beams the spatial overlap with the pump mode is only partial (see algorithm for exact dependence). Thus, as the pulse energy increases, gain saturation pushes the beam towards wider beam waists to fully extract the pump energy. Finally, however, diffraction losses come into effect, as shown in the \textbf{right axis} - Diffraction transmission (1-loss) as a function of beam waist (green curve). The beam waist at steady state would correspond to the maximal net gain - $g \cdot l_D$, which occurs near the equality point (marked by the black circle).}
    \label{fig:gain}
\end{figure}

The nonlinear diffraction losses due to the Kerr-lensing effect are at the heart of KLM. In the single-mode regime, the Kerr-lensing effect couples the beam size and the instantaneous intensity at the crystal. Thus, by inserting a mechanism that penalizes the laser for large beams \cite{parshani_diffractive_2021,yefet_review_2013,wright_mechanisms_2020}, we can create an effective saturable absorber. Here we note two competing loss/saturation mechanisms - small beams suffer from insufficient overlap with the pump mode, which reduces their gain, whereas larger beams suffer increased loss due to the effective Kerr lens aperture. This competition is summarized in figure \ref{fig:gain}. In general, the nonlinear losses can be divided into two-categories: hard apertures, where any power beyond a certain beam size is completely cut off; and soft apertures, where beams beyond a certain size incur a certain amount of loss that depends on the beam's size \cite{Juang1997, yoo_byung_duk_numerical_2005, t_brabec_hard-aperture_1993}. Both can be easily incorporated into our simulation.

\begin{figure}
    \centering
    \includegraphics[width = 0.7\textwidth]{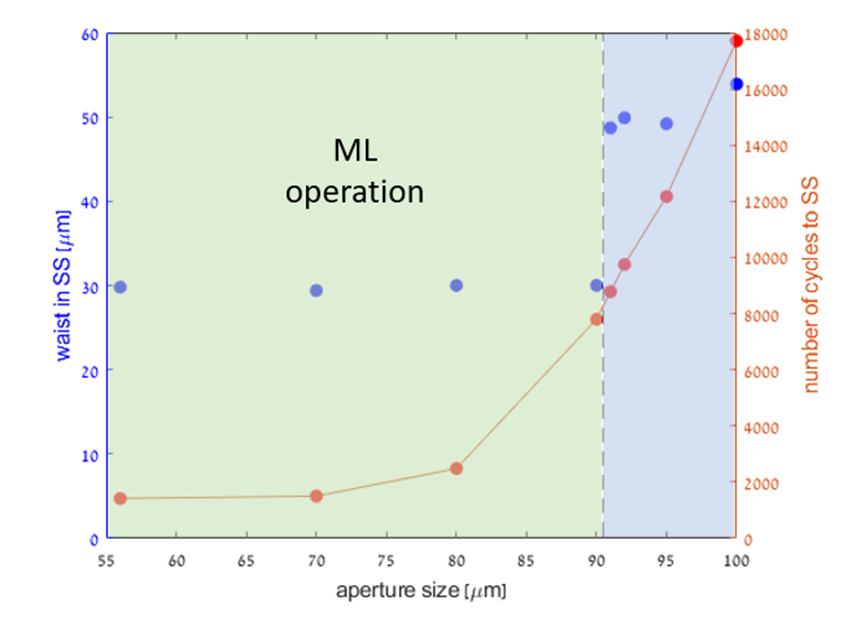}
    \caption{Artificial aperture for numerical convergence. In order to maintain a reasonable beam size during all stages of the simulation we implemented an artificial aperture to prevent the beam size from diverging during the initial stage of cavity evolution, while the intra-cavity power is too low to support a significant Kerr lens. This aperture was implemented as a lens with an imaginary focus, which is equivalent to a Gaussian aperture that limits the beam width, but keeps its total energy. To verify that the aperture does not affect the ML results we ran the simulation for various artificial aperture sizes, recording the beam waist at steady state \textbf{(left axis)}, as well as the number of round trips required to reach steady state \textbf{(right axis)}. Note that within the convergence range (green background) the actual beam waist in steady-state is not affected by the artificial aperture. The run-time, however, diverges for wider aperture sizes until the algorithm stops to converge at all.}
    \label{fig:aperture_vs_runtime}
\end{figure}

In our simulation, we opt for a soft-aperture mechanism, corresponding to a situation where the most important loss mechanism is the spatial overlap between the laser and the pump mode at the gain medium. Different mechanisms simply amount to different loss functions, and can be easily accommodated by our software. For numerical reasons, in order to assure convergence of the cavity evolution (power and beam parameter) to the steady state, the beam size must be limited by an artificial hard aperture, which prevents the beam size from diverging during the initial stage of cavity evolution, while the intra-cavity power is too low to support a significant Kerr lens. This does not affect the dynamics of the cavity or the steady-state value, but only the convergence time of the simulation, as illustrated in figure \ref{fig:aperture_vs_runtime}. 

The simulation is performed partially in time-domain, where the instantaneous nonlinear effects are calculated, and partially in frequency domain where the linear transfer function of the cavity is calculated detailed in the pseudo-code given in algorithm \ref{code:main}. 
To propagate the field in both time and space, the simulation is divided into a spatial procedure, and a temporal procedure. The spatial procedure, detailed in the pseudo-code given in algorithm \ref{code:spatial}, handles the spatial ABCD propagation under the Gaussian single-mode assumption (including the spatial Kerr and the diffraction losses). The temporal procedure, detailed in the pseudo-code given in algorithm \ref{code:time}, handles the temporal evolution of SPM, gain, loss and dispersion. The two components are coupled through the Kerr-lens interaction, where the temporal profile of the beam determines the optical power of the Kerr-lens, which in turn affects the temporal evolution via the diffraction losses. The variables and function names are given in table \ref{table:sim_vars}.

\begin{table}
\label{table:sim_vars}
    \centering
    \begin{tabular}{c|l}
        \textbf{Simulation Parameter} & \textbf{    \space \space \space Description} \\
        \hline
        $N$ & Total number of cavity round-trips in the simulation. \\
        $M$ & Total number of time-points within each round-trip. \\
        $w(t)$ & Instantaneous beam waist, $M$-vector for each round. \\
        $R(t)$ & Beam radius of curvature, $M$-vector.\\
        $P(t)$ & Beam power, $M$-vector.\\
        $I(t)$ & Beam intensity at the waist, $M$-vector. \\
        $q(t)$ & Complex beam radius, $M$-vector. \\
        $g_0$ & Small-signal gain. \\
        $g(\bar{P},w(t))$ & Saturated gain (depends on waist and average power). \\
        $l$ & Cavity loss rate. \\
        $\sigma_G$ & Gain bandwidth (assuming Lorentzian spectrum). \\
        $E(t)$ & Field in time, $M$-vector. \\
        $\hat{E}$ & Field in frequency, $M$-vector. \\
        $\sigma_D$ & Linear dispersion (GVD). \\
        $L_1$ & Right cavity arm-length. \\
        $L_2$ & Left cavity arm-length.\\
        $L_c$ & Crystal's length.

    \end{tabular}
    \caption{Names and description of the variables used in the algorithms.}
\end{table}

\begin{algorithm}[H]
    \caption{Main loop with temporal Kerr-lens}
    \label{code:main}
    \begin{algorithmic}
    \Procedure{Main}{$N, M$}
        \State \textbf{Initialization:}
        \State $w_p \gets 30\mu m$ \Comment{Initialize pump beam waist (for mode matching).}
        \State $w[:] \gets \textit{Random}$ \Comment{Initialize temporal beam waist $w(t)$ from a random distribution.}
        \State $R[:] \gets \textit{Random}$ \Comment{Initialize beam radius of curvature (phase) $R(t)$ from a random distribution.}
        \State $\frac{1}{q[:]} \gets \frac{1}{R[:]} - \frac{\lambda i}{\pi w[:]^2}$  \Comment{Complex beam parameter definition.} 
        \State $E[:] \gets \textit{Random}$     \Comment{Initialize temporal field amplitude to noise.}
        \State $s(\omega) \gets \frac{1}{1 + \left(\frac{\omega}{\sigma_G}\right)^2}$ \Comment{Spectral shape of the linear gain/loss (Lorentzian spectrum).}
        \State $\phi(\omega) \gets \sigma_D^2 \omega^2$ \Comment{Spectral phase per round trip due to dispersion (2nd order GVD).}
            
        \For{$i \gets 1$ to $N$}    \Comment{ Main loop of $N$ round trips}
            \State $\hat{E}[:] \gets \text{FFT}(E[:])$ \Comment{Transforming to frequency domain to calculate linear propagation.}
           \State $\hat{E}[:] \gets (l \cdot s(\omega) \cdot e^{-i\phi(\omega)}) \hat{E}[:]$ \Comment{apply linear loss, spectral gain shaping and chromatic dispersion.}
            \State $E[:] \gets \text{IFFT}(\hat{E}[:])$ \Comment{Transform field back to time domain}
            \State $P[:] \gets {|E[:]|}^2 \pi w^2[:]$
            \Comment{Calculate the instantaneous power $P(t)$}

            \State $q[:],\phi_{NL}[:] \gets  \text{ML-Spatial}[\delta, P[:], q[:], \gamma, L_1, L_2, L_c]$ \Comment{Procedure to calculate the new beam parameter $q(t)$ and the nonlinear phase $\phi_{NL}(t)$ for the next round trip.}
             \State $g[:] \gets  \text{SaturateGain}[P[:],q[:],w_{p},U_{sat}]$ \Comment{Procedure to calculate the gain saturation based on power $P(t)$ and the temporal overlap with pump mode.}
            \State $l_D[:] \gets  \text{DiffLoss}[q[:],w_p]$ \Comment{Procedure to calculate the diffraction losses based on beam waist relative to the pump.}
            \State $E[:] \gets  l_D[:] \cdot g[:] \cdot e^{i\phi_{NL}[:]}E[:]$ 
            \Comment{Apply saturated gain, diffraction losses and nonlinear phase to the field in time}
        \EndFor 
    \EndProcedure
    \end{algorithmic}
\end{algorithm}

\begin{algorithm}[H]
    \caption{\bf{\space\space Procedures for cavity propagation with spatial Kerr-lens}}
    \label{code:spatial}
    \begin{algorithmic}
        \Procedure{ML-Spatial}{$\delta, P, q, w, \gamma, L_1, L_2, L_c$} 
            \Comment{Cavity propagation with time dependence}
            \State $M_{prop}(z) \gets \begin{pmatrix}
            1 & z \\ 0 & 1 \end{pmatrix}$
            \Comment{matrix for spatial propagation} 
            \State $M_{KL}(f_{K}, f_{A}) \gets \begin{pmatrix}
            1 & 0 \\ -\frac{1}{f_K}-i\frac{1}{f_A} & 1 \end{pmatrix}$
            \Comment{Non-linear lens matrix (K) with artificial aperture (A)} 
            \State $f_{A} \gets \frac{2\pi {w^2_{aperture}}}{\lambda} $
            \Comment{Artificial aperture as a lens with imaginary focus} 
           
            \State $q[:] \gets \text{PropagateThickKerrLens}[q[:],P[:]]$ \Comment{calculate $q(t)$ at the output of Kerr medium ($1^{\text{st}}$ pass)}
            \State $M =  M_{Left}$,\space\space $q[:] \gets \frac{(A q[:]+B)}{(C q[:]+D)}$ \Comment{linear propagation through left side of cavity}

            \State $q[:] \gets \text{PropagateThickKerrLens}[q[:],P[:]]$ \Comment{calculate $q(t)$ at output of Kerr medium ($2^{\text{nd}}$ pass)}
            
            \State $M =  M_{Right}$,\space\space $q[:] \gets \frac{(A q[:]+B)}{(C q[:]+D)}$ \Comment{linear propagation through right side of cavity}
         \EndProcedure

        \Procedure{PropagateThickKerrLens}{$P(t), q(t)$}  \Comment{the thick Kerr medium of length $L$ is broken into several thin Kerr lenses separated by short free propagation $l$ } 
            \State $M[:] \gets M_{prop}(l/2),$ \space \space $q[:] \gets \frac{(A q[:]+B)}{(C q[:]+D)}$ \Comment{The propagation matrix of the current step in the cavity (we start at the entrance to the nonlinear medium)} 
            \For{$j \gets 1$ \textbf{to} L/l}
            \Comment{propagation through the thin Kerr lenses separated by free propagation $l$ }
                \State $f_K[:],\phi_{NL}[:]\gets \text{CalculateThinKLMFocusAndPhase}[w(t), P(t)]$
                \Comment{Calculation of the nonlinear focus and phase for each thin Kerr lens}
                \State {$M[:] \gets M_{prop}(l) \cdot M_{KL}^{j}[:] \cdot M[:],$ \space \space $q[:] \gets \frac{(A q[:]+B)}{(C q[:]+D)}$}
            \EndFor
        \EndProcedure
       
         \Procedure{CalculateThinKLMFocusAndPhase}{$P(t), q(t)$} 
            \Comment{Calculate the nonlinear focus $f_K(t)$ and phase $\phi_{NL}(t)$ based on the temporal power and beam waist}
            \State{$w[:] \gets \sqrt{\frac{\pi}{\lambda}\text{Im}\frac{1}{q[:]}}$}
            \Comment{Calculate the instantaneous beam waist}
            \State{$f_{K}[:] \gets \frac{\pi}{8n_{2}l} \frac{w^4[:]}{ \cdot P[:]}$}
            \Comment{Non-linear focus}
            \State $\phi_{NL}[:] \gets \frac{4n_{2}l}{\pi \lambda} \frac{P[:]}{ w^2[:]}$
            \Comment{Nonlinear self-phase modulation}
            \State{$\mathbf{return}$ \space $\phi_{NL}[:],f_{K}[:]$ }
        \EndProcedure
    \end{algorithmic}
\end{algorithm}

\begin{algorithm}[H]
    \begin{algorithmic}
        \Procedure{SaturateGain}{$P(t),W(t),W_{p},U_{sat}$} 
            \State{$U \gets  \mathbf{Sum}(P[:])$}
            \Comment{Total pulse energy of the round trip (for gain saturation)}
            \For{$i \gets 1$ to $M$}
                \If{$w[i] < w_p$}
                    \State $ g[i] = \frac{g_0}{1 + \frac{U}{U_{sat}\cdot \left(\frac{w[i]}{W_{p}}\right)^2}}$
                \Else 
                    \State $ g[i] = \frac{g_0}{1 + \frac{U}{U_{sat}}}$ 
                \EndIf{}
            \EndFor
        \State{$\mathbf{return}$\space $g[:]$}
        \EndProcedure
        \\
       \Procedure{DiffLoss}{$w(t),w_{p}$} 
        \Comment{Diffraction loss calculation}
            \For{$i \gets 1$ to $M$}
                \If{$w[i] > w_p$}
                \Comment{Mode matching losses for spot size wider than the pump spot size}
                    \State{$l_D[i] = \frac{1}{(1 + {(w[i]-w_{p})}^2/(2 w_{p}^2))}$ }
                \Else 
                    \State{$l_D[i] = 1$}
                \EndIf{}
            \EndFor
        \State{$\mathbf{return}$ \space $l_D[:]$}
        \EndProcedure 
    \end{algorithmic}
    \caption{\bf{\space\space Procedures for non-linear focus and gain saturation.}}
\label{code:time}
\end{algorithm}

\section{Conclusions}
    \label{sec:conclusions}
In this paper, we introduced a complete algorithm, accompanied by an open-source MATLAB package for simulating the real-time dynamics of a Kerr-lens mode-locked laser. We simulate from basic principles the complete coupled spatio-temporal dynamics of the laser cavity, under only the assumption of a single Gaussian spatial mode, allowing to observe the complete dynamical evolution of the pulse field in the cavity from the initial noise field to the final steady state. The simulation was thoroughly validated by reproducing all the basic properties of Kerr-lens mode-locked lasers, such as the dependence of the pulse duration in the gain bandwidth, the dispersion and the self-phase modulation with quantitative agreement to known experimental values. 
We demonstrate the evolution of pulse energy and duration, as well as the gain dynamics that determine them. Furthermore, the simulation also predicted novel phenomena of space-time coupling in linear KLM cavities \cite{parshani_diffractive_2021,Parshani2021}, which demonstrates the power of the simulation as a tool for the design and analysis of ultrafast lasers.


\bibliography{main.bib}
\end{document}